\documentclass[%
superscriptaddress,
showpacs,
longbibliography,
amsmath,amssymb,
aps,
twocolumn
]{revtex4-1}
\usepackage{bbm}
\usepackage{graphicx}
\usepackage{dcolumn}
\usepackage{bm}
\usepackage{natbib}
\usepackage{times}
\usepackage{units}
\usepackage{color}
\usepackage[pdftex, colorlinks=true, linkcolor=blue, citecolor=blue,
urlcolor=blue]{hyperref}
\usepackage{comment}
\usepackage{textcomp}

\newcommand{\bs}[1]{\boldsymbol{#1}} 
\newcommand{\bsl}[1]{\mathbf{#1}} 

\definecolor{smaragd}{cmyk}{.9,0.3,.9,.1}

\begin{document}

\title{Plasmon polaritons in cubic lattices of spherical metallic nanoparticles}

\author{Simon Lamowski}
\affiliation{Department of Physics and Center for Applied Photonics,
  University of Konstanz, D-78457 Konstanz, Germany}
\affiliation{Okinawa Institute of Science and Technology Graduate University, Onna-son, Okinawa 904-0395, Japan}
\author{Charlie-Ray Mann}
\affiliation{School of Physics and Astronomy, University of Exeter, Stocker Rd., Exeter EX4 4QL, United Kingdom}
\author{Felicitas Hellbach}
\affiliation{Department of Physics and Center for Applied Photonics, University of Konstanz, D-78457 Konstanz, Germany}
\author{Eros Mariani}
\affiliation{School of Physics and Astronomy, University of Exeter, Stocker Rd., Exeter EX4 4QL, United Kingdom}
\author{Guillaume Weick}
\email{guillaume.weick@ipcms.unistra.fr}
\affiliation{Universit\'e de Strasbourg, CNRS, Institut de Physique et Chimie des Mat\'eriaux de Strasbourg, UMR 7504, F-67000 Strasbourg, France}
\author{Fabian Pauly}
\email{fabian.pauly@oist.jp}
\affiliation{Department of Physics and Center for Applied Photonics,
  University of Konstanz, D-78457 Konstanz, Germany}
\affiliation{Okinawa Institute of Science and Technology Graduate University, Onna-son, Okinawa 904-0395, Japan}

\begin{abstract}
We theoretically investigate plasmon polaritons in cubic lattices of spherical
metallic nanoparticles.  The nanoparticles, each supporting triply-degenerate
localized surface plasmons, couple through the Coulomb dipole-dipole
interaction, giving rise to collective plasmons that extend over the whole
metamaterial. The latter hybridize with photons forming plasmon polaritons,
which are the hybrid light-matter eigenmodes of the system. We derive general
analytical expressions to evaluate both plasmon and plasmon-polariton
dispersions, and the corresponding eigenstates. These are obtained within a
Hamiltonian formalism, which takes into account retardation effects in the
dipolar interaction between the nanoparticles and considers the dielectric
properties of the nanoparticles as well as their surrounding. Within this
model we predict polaritonic splittings in the near-infrared to the visible
range of the electromagnetic spectrum that depend on polarization, lattice
symmetry and wavevector direction.  Finally, we show that the predictions of
our model are in excellent quantitative agreement with conventional
finite-difference frequency-domain simulations, but with the advantages
of analytical insight and significantly reduced computational cost.
\end{abstract}

\maketitle

\section{Introduction}
Plasmonic metamaterials can be exploited to manipulate light at subwavelength
scales and may be used to tailor optical properties \cite{Barnes2003, maier,
  Halas2011}. They consist of meta-atoms with possibly complicated
subwavelength structures that are arranged in a controlled fashion
\cite{Meinzer2014}. Potential applications of such metamaterials range from
optical cloaking over planar hyperlenses to optical data processing
\cite{Shalaev2008,Tame2013}.

The study of the optical properties of one-dimensional (1D),
  two-dimensional (2D) and three-dimensional (3D) arrays of metallic particles
  is a very active field of research \cite{Ross2016}. In the past, most
  theoretical and experimental research has been focused on 1D and 2D systems,
  since they are much easier to fabricate with well-established techniques
  \cite{Halas2011,Meinzer2014}. However, the development of reliable
  techniques to control 3D assemblies of plasmonic nanoparticles is presently
  making substantial advances, and such 3D assemblies can now be achieved by
  using surface ligands or DNA templates
  \cite{Tan2011,Kim2016,Liu2016,Ross2016,Malassis2016}. It is thus of current
  interest to also understand systematically the structure-property
  relationships in 3D crystalline arrangements of meta-atoms, where, beside
  the shape and the size of the nanoparticles themselves, the spacing and the
  crystal symmetry can be controlled independently.

The optical properties of a plasmonic metamaterial are governed in the
  first instance by those of the individual metallic nanoparticles
  \cite{Meinzer2014}.  Of primary importance to understand such optical
properties are the localized surface plasmons (LSPs), which correspond to
collective oscillations of the valence electrons against the ionic background.
The resonance frequency and polarization of the LSP modes are determined by
the size, shape and material of the nanoparticles.

Classical electrodynamics can be used to understand many of the optical
  properties of 1D, 2D, and 3D plasmonic metamaterials \cite{maier,
    Ross2016}. Depending on the distance between the meta-atoms, two
  qualitatively different regimes emerge \cite{Meinzer2014}: In the first
  regime, the distance between the meta-atoms is on the order of or larger
  than the wavelength associated with the LSP resonance of individual
  nanoparticles, so that diffractive far-field interactions between the
  meta-atoms of the array can interfere, leading to collective modes termed
  surface lattice resonances. In the second, opposite regime, the meta-atom
  separation is much smaller than the LSP resonance wavelength so that
  near-field interactions are predominant, yielding collective plasmons that
  are extended over the whole metamaterial. In the present work we concentrate
  on the latter regime.

Early studies on the plasmonic properties of near-field-coupled metallic
nanoparticles focused on 1D chains using a nonretarded model of point dipoles
\cite{Quinten1998,approx1,maier03_PRB,approx2}, followed by fully-retarded
classical approaches applied to 1D
\cite{weber04_PRB,citri04_NL,Simovski2005,citri06_OL,koend06_PRB,marke07_PRB,Fung2007,koend09_NL,Samrowski2010,petro15_PRA}
and 2D systems \cite{Moreno2002,Hafner2005,Garcia-Martin2005, Sannomiya2011,
  Markos2016}. Three-dimensional metastructures were also investigated using
more approximate approaches such as the Maxwell-Garnett effective medium
theory \cite{Ross2016} or Bruggeman effective medium theory
\cite{Santiago2017}.  In addition to the classical, typically fully numerical
treatments, an analytically tractable approach based on a Hamiltonian
formalism was recently applied to 1D \cite{lee12_PRA, brand16_PRB,
  Downing2017,Downing2018}, 2D \cite{Weick2013, Sturges2015, Mann2017} and 3D systems
\cite{Weick}.

In this work we study the less explored 3D plasmonic arrays in the regime of
near-field coupling between spherical metallic nanoparticles. Spherical
particles are chosen in order to focus on the effects of crystal structure on
the optical properties only. The nature of the modes supported by a plasmonic
metamaterial depends crucially on the dimensionality of the lattice. For 1D
and 2D lattices, the collective plasmons couple to a continuum of photonic
modes with different wavevector components along directions where
translational symmetry is absent. However, as it has been pointed out by
  Hopfield in the context of exciton polaritons \cite{hopfield}, in stark
  contrast to lower dimensional systems, collective plasmons in 3D lattices
  only couple to photons which conserve crystal momentum due to the discrete 
  translational symmetry of the system. As a result, the true eigenmodes of the 
  metamaterial are coherent superpositions of plasmons and photons, which we call plasmon polaritons. 
  We study them by means of an analytically tractable
  Hamiltonian-based approach, which importantly incorporates retardation
  effects.
  
In what follows we consider 3D lattices of spherical metallic nanoparticles,
including simple cubic (sc), face-centered cubic (fcc), and body-centered
  cubic (bcc) structures. In the quasistatic limit \cite{kreibig}, each
nanoparticle supports a discrete set of multipolar LSP modes. However, as we
consider small nanoparticles (of some $\unit[10]{nm}$ in radius), we neglect
higher-order multipolar modes and focus on the fundamental dipolar LSPs, whose
corresponding frequency lies in the visible to ultraviolet range of the
spectrum.  In this regime, quantum-size effects in the optical
response of the nanoparticles can be significant \cite{Tame2013}.  Due to the
spherical symmetry of the nanoparticles, each dipolar LSP is triply-degenerate
with three polarization degrees of freedom.

We work in the
  Coulomb gauge \cite{cohen, craig}, where the scalar and vector potentials
  describe the longitudinal and transverse components of the electromagnetic
  field, respectively. The scalar potential, which depends only on the matter
  degrees of freedom, takes the form of the instantaneous Coulomb interaction
  between the LSPs. This results in collective plasmonic modes, which extend across the
  whole metamaterial. The effects of retardation are then included in the
  light-matter coupling through the interaction of the LSPs with the
  transverse vector potential. In this way, transverse photons hybridize with the collective plasmons to form plasmon polaritons. We also take into account screening effects from the core electrons as well as the dielectric medium surrounding the nanoparticles.

Here, we decisively
  extend inspiring work of some of the authors~\cite{Weick}. Although it is
  stated in Ref.~\cite{Weick} that spherical metallic nanoparticles are used,
  these nanoparticles were assumed to exhibit only one polarization degree of
  freedom that was fixed in a given direction. This gives rise to a single
  plasmon band, whose polarization does not depend on the wavevector. In fact
  this model does not correctly describe lattices of spherical nanoparticles,
  but could be used to study lattices of resonators that have a nondegenerate
  fundamental eigenmode, such as plasmonic nanorods. Our treatment fixes this issue by considering plasmon
  polaritons which arise from the hybridization of photons with three
  plasmonic bands with wavevector-dependent polarizations. Furthermore we show
  that the model yields plasmon-polariton properties in excellent agreement
  with classical electrodynamics simulations at a much reduced computational
  cost and at the benefit of analytical intuition. With our newly developed
  tools, we demonstrate that these highly symmetric cubic systems exhibit
  polarization-dependent optical properties such as band splittings in the
  near-infrared or visible range of the spectrum. With the emerging
  fabrication techniques for 3D metallic nanoparticle lattices, this work is
  an important step towards accurate predictions of their polaritonic
  properties, and the model can be readily extended to more complex lattices
  and nanoparticle shapes.

The paper is organized as follows: In Sec.~\ref{sec:model} we describe our
theoretical model to study plasmon polaritons.  The general solution to this
model is subsequently presented in Sec.~\ref{sec:PP}. The resulting dispersion
relations of the collective plasmons and plasmon polaritons for sc, fcc and
bcc lattices are discussed in Secs.~\ref{sec:CP} and \ref{sec:full},
respectively. In Sec.~\ref{sec:compare}, we compare our predictions to
classical electrodynamics simulations. We finally summarize our results in
Sec.~\ref{sec:ccl}. In the Appendix we discuss the form of the dielectric
tensor that shows a nonlocal response.

\section{Model}
\label{sec:model}

\begin{figure}[tb]
\includegraphics[keepaspectratio=true, width=\columnwidth ]{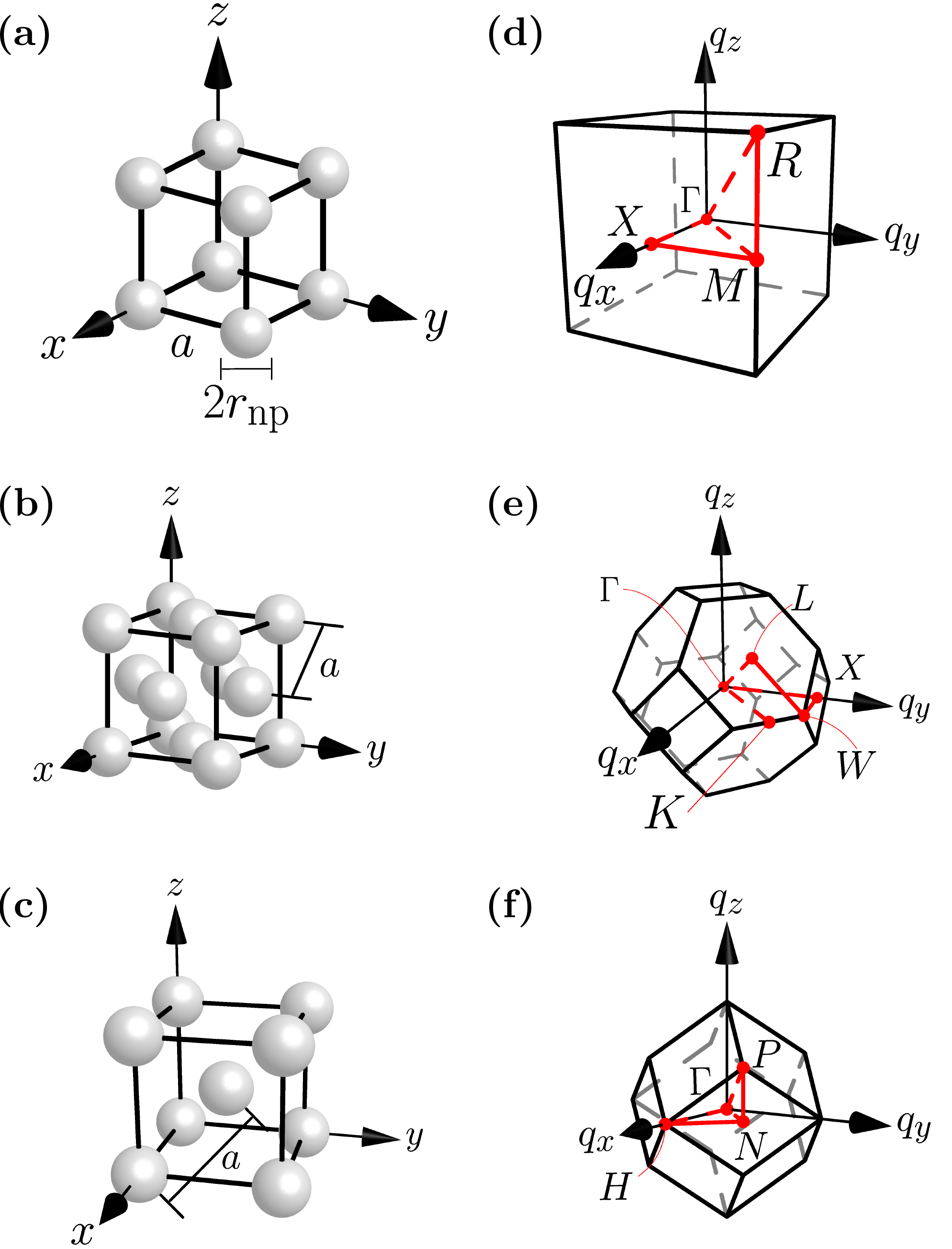}
\caption{Conventional unit cells for (a) sc, (b) fcc, and (c) bcc lattices of
  spherical metallic nanoparticles of radius $r_{\text{np}}$ with the
  primitive lattice parameter $a$. (d)-(f) Corresponding first Brillouin
  zones, where the red lines indicate the paths, over which the plasmon and
  plasmon-polariton dispersions are plotted in
  Figs.~\ref{pl-dis}-\ref{fig:QM+Comsol}.}
\label{fig:latbzs}
\end{figure}

We consider sc, fcc, and bcc lattices of spherical metallic nanoparticles
separated by a center-to-center distance $a$ between nearest neighbors, as
depicted in Figs.~\ref{fig:latbzs}(a)-(c). The corresponding first Brillouin
zones are shown in Figs.~\ref{fig:latbzs}(d)-(f). We describe the nanoparticles
with a Drude-like dielectric function
\begin{equation}
 \epsilon_{\mathrm{r}}^{\mathrm{D}}(\omega) = \epsilon_{\mathrm{d}} -
 \frac{\omega_{\mathrm{p}}^2 }{\omega(\omega+\mathrm{i} \gamma^{\mathrm{D}})},\label{eq:epsDrude}
\end{equation}
where $\omega_{\mathrm{p}}$ is the plasma frequency of the considered (noble)
metal and where the dielectric constant $\epsilon_{\mathrm{d}}$ takes into
account the screening of the conduction electrons by the $d$-electrons. In
our model we use $\gamma^{\mathrm{D}}=0$, but we will explore the effect of a
nonvanishing Drude damping in the finite-difference frequency-domain (FDFD)
calculations, presented in Sec.~\ref{sec:compare}. The surrounding medium that
fills the space between the nanoparticles is characterized by the dielectric
constant $\epsilon_{\mathrm{m}}$. The magnetic permeabilities of the
nanoparticles and the embedding medium are assumed to be equal to the vacuum
permeability. Each nanoparticle in the lattice supports three degenerate
dipolar LSPs polarized in the $x$, $y$ or $z$ direction. They interact with
their neighbors through the quasistatic dipole-dipole interaction
\begin{equation}
  \label{eq:dd}
  V_\mathrm{dip}(\mathbf{R}, \mathbf{R}')=\frac{9\epsilon_{\mathrm{m}}
  }{\left(\epsilon_{\mathrm{d}}+2\epsilon_{\mathrm{m}} \right)^2}
  \frac{\mathbf{p}\cdot\mathbf{p}'-3(\mathbf{p}\cdot\hat n)(\mathbf{p}'\cdot\hat
    n)}{4\pi\epsilon_0|\mathbf{R}-\mathbf{R}'|^3},
\end{equation}
where $\mathbf{p}$ and $\mathbf{p}'$ are the dipole moments associated with
the LSPs of the nanoparticles located at the lattice sites $\mathbf{R}$ and
$\mathbf{R}'$, respectively, while $\hat
n=(\mathbf{R}-\mathbf{R}')/|\mathbf{R}-\mathbf{R}'|$, and $\epsilon_0$ is the
vacuum permittivity. Here and in what follows, hats denote unit vectors. In
the expression above, the prefactor takes into account the two dielectric
environments and arises from a model in which each point dipole is located
inside a sphere with dielectric constant $\epsilon_\mathrm{d}$, and separated
by a medium with dielectric constant $\epsilon_\mathrm{m}$ \cite{Riande2004}.
As we only consider dynamical degrees of freedom relating to the fundamental
dipolar LSPs, we are thus neglecting any effects of higher-order multipolar
plasmons. This approximation has been shown to be valid for center-to-center
interparticle separations $a\gtrsim 3r_{\mathrm{np}}$ \cite{approx1}, with
$r_{\mathrm{np}}$ the nanoparticle radius (see Fig.~\ref{fig:latbzs}). We
demonstrate the validity of this approximation in Sec.~\ref{sec:compare} by
comparing our results to FDFD simulations.

We write the full
Hamiltonian of the system as
\begin{equation}
   H =H_{\mathrm{pl} } + H_{\mathrm{ph} } + H_{\mathrm{pl}\textrm{-}\mathrm{ph} },
   \label{eq:H_full}
\end{equation}
where $H_{\mathrm{pl}}$ and $H_{\mathrm{ph}}$ denote the plasmonic and
photonic Hamiltonians, respectively, and where
$H_{\mathrm{pl}\textrm{-}\mathrm{ph} }$ is the interaction Hamiltonian between
both subsystems.  In the Coulomb gauge \cite{cohen, craig}, the purely
plasmonic Hamiltonian reads~\cite{Weick2013, Sturges2015, Weick,
  brand16_PRB, Downing2017}
\begin{align}
 H_{\mathrm{pl}} =&\;\hbar\omega_0 \sum\limits_{\bsl{q},\hat\sigma}
 b_{\bsl{q}}^{\hat\sigma\dagger} b_{\bsl{q}}^{\hat\sigma}
 \nonumber\\
   &
   +\hbar\Omega \sum\limits_{\bsl{q},\hat\sigma,\hat\sigma'}
   f_{\bsl{q}}^{\hat\sigma,\hat\sigma'}\left[b_{\bsl{q}}^{\hat\sigma\dagger}\left(b_{\bsl{q}}^{\hat\sigma'}+b_{\bsl{-q}}^{\hat\sigma'\dagger}\right)+ \mathrm{h.c.} \right],
   \label{eq:H_pl}
\end{align}
with
\begin{equation}
  f_{\bsl{q}}^{\hat\sigma,\hat\sigma'} =
  \sum_{\substack{\bs{\rho}\\(a\leqslant\rho\leqslant\rho_\mathrm{c})}}
  \left(\frac{a}{\rho}\right)^3  \frac{\cos{\left(\bsl{q}\cdot\bs{\rho} \right)}}{2}
  \left[\delta_{\hat\sigma\hat\sigma'}-3(\hat\sigma\cdot\hat{\rho})(\hat\sigma'\cdot\hat{\rho})\right].
  \label{fq_mat}
\end{equation}
Here, $\bsl{q}=q\, \hat q$ is the plasmonic wavevector in the first Brillouin zone. In
Eq.~\eqref{eq:H_pl}, $b_{\bsl{q}}^{\hat\sigma}=\mathcal{N}^{-1/2}
\sum_\mathbf{R}\exp{(-\mathrm{i}\mathbf{q}\cdot\mathbf{R})}b_\mathbf{R}^{\hat\sigma}$
is defined as the Fourier transform of the bosonic operator
$b_\mathbf{R}^{\hat\sigma}$, which annihilates an LSP at lattice site
$\mathbf{R}$ with polarization $\hat\sigma=\hat x$, $\hat y$ or $\hat z$,
where $\mathcal{N}$ is the number of unit cells of the metacrystal.  The first
term on the right-hand side of Eq.~\eqref{eq:H_pl} describes the uncoupled
LSPs with Mie frequency \cite{kreibig}
\begin{equation}
  \omega_0=\frac{\omega_{\mathrm{p}}}{\sqrt{\epsilon_{\mathrm{d}}+2\epsilon_{\mathrm{m}}}},\label{eq:omegaLSP}
\end{equation}
while the second one with coupling constant
\begin{equation}
  \Omega=\frac{3\epsilon_{\mathrm{m}}}{2(\epsilon_{\mathrm{d}}+2\epsilon_{\mathrm{m}})}\omega_0
\left(\frac{r_\mathrm{np}}{a}\right)^3
\label{eq:Omega}
\end{equation}
corresponds to the Coulomb dipole-dipole interaction [cf.\ Eq.~\eqref{eq:dd}]
between nanoparticles linked by the separation vector $\boldsymbol\rho$.
Crucially, we consider Coulomb interactions up to a large cut-off distance
$\rho_\mathrm{c}\gg a$, beyond the nearest-neighbor approximation that was
employed in Ref.\ \cite{Weick}. As will be highlighted later, these long-range
Coulomb interactions are critical for obtaining the correct plasmonic
dispersions.

As discussed in detail in Ref.~\cite{Cohen1955},
there is a region of slow convergence of
$f_{\bsl{q}}^{\hat\sigma,\hat\sigma'}$  around the
$\Gamma$ point [see Eq.~\eqref{fq_mat}]. This stems from discontinuities of
$f_{\bsl{q}}^{\hat\sigma,\hat\sigma'}$ at $\mathbf{q}=0$ for $\rho_\mathrm{c}
\rightarrow \infty$. These discontinuities lead to the Gibbs-Wilbraham
phenomenon \cite{Hewitt1979}, and the summation in Eq.~\eqref{fq_mat} does not
easily converge with increasing cutoff radius $\rho_\mathrm{c}$. Thus, for
small wavevectors $q< \alpha\rho_\mathrm{c}^{-1}$, with $\alpha$ a real
positive number, we use the correction $f_{\bsl{q}}^{\hat\sigma,\hat\sigma'}
=-2\pi\left[\delta_{\hat\sigma\hat\sigma'}-3(\hat\sigma\cdot\hat{q})(\hat\sigma'\cdot\hat{q})
  \right]/3\nu$ for the infinite lattice \cite{Cohen1955}. It contains the
factor $\nu$, which accounts for the different volumes of the
primitive cells of the considered lattices and equals $\nu=1$ for sc,
$\nu=2^{-1/2}\simeq0.71$ for fcc, and $\nu=4/3^{3/2}\simeq0.77$ for bcc
lattices, respectively.

In Eq. \eqref{eq:H_full} the photonic subsystem is described by
\begin{equation}
   H_{\mathrm{ph} }=\sum\limits_{\bsl{q},\hat\lambda_{\bsl{q}}
   }\hbar\omega_{\mathrm{ph},\bsl{q}}c_{\mathbf{q}}^{\hat\lambda_{\bsl{q}}\dagger}c_{\bsl{q}}^{\hat\lambda_{\bsl{q}}
   }, \label{eq:H_ph}
\end{equation}
where $c_{\bsl{q}}^{\hat \lambda_{\bsl{q}}}$ annihilates and
$c_{\mathbf{q}}^{\hat\lambda_{\bsl{q}}\dagger}$ creates a photon with
wavevector $\bsl{q}$, dispersion
$\omega_{\mathrm{ph},\bsl{q}}=cq/\sqrt{\epsilon_{\mathrm{m}}}$, and transverse
polarization $\hat{{\lambda}}_{\bsl{q}}$ (with
$\hat{{\lambda}}_{\bsl{q}}\cdot\mathbf{q}=0$). Here
$c/\sqrt{\epsilon_{\mathrm{m}}}$ is the speed of light in the embedding
medium. In the long-wavelength limit $qr_\mathrm{np}\ll 1$, the minimal
light-matter coupling Hamiltonian in Eq.~\eqref{eq:H_full} takes the form
\begin{align}
\label{eq:H_pl-ph}
   H_{\mathrm{pl}\textrm{-}\mathrm{ph} }=&\; \mathrm{i}\,\hbar\omega_0 \sum_{\bsl{q},\hat\sigma,\hat\lambda_{\bsl{q}}}\hat{\sigma}\cdot{\hat{\lambda}}_{\bsl{q}} \xi_{\bsl{q}}
      \left(
      b_{\bsl{q}}^{\hat\sigma\dagger}c_{\bsl{q}}^{\hat\lambda_{\bsl{q}}}
      + b_{\bsl{q}}^{\hat\sigma\dagger}c_{-\bsl{q}}^{\hat\lambda_{\bsl{q}}\dagger}
      -\mathrm{h.c.}\right)
      \nonumber\\
      &+\hbar\omega_0\sum_{\bsl{q},\hat\lambda_{\bsl{q}}}\xi_{\bsl{q}}^2
      \left(c_{\bsl{q}}^{\hat\lambda_{\bsl{q}}\dagger}c_{\bsl{q}}^{\hat\lambda_{\bsl{q}}}
      +c_{\bsl{q}}^{\hat\lambda_{\bsl{q}}\dagger}c_{-\bsl{q}}^{\hat\lambda_{\bsl{q}}\dagger} + \mathrm{h.c.} \right),
\end{align}
 where $\xi_{\bsl{q}}=[2\Omega\pi/(\nu\omega_{\mathrm{ph},\bsl{q}})]^{1/2}$.
 Since we consider lattice constants $a$ much smaller than the wavelength
 associated with the LSP resonances, we neglect Umklapp processes in
   Eqs.~(\ref{eq:H_ph}) and (\ref{eq:H_pl-ph}).  However, the model can be
 readily extended to include such Umklapp scattering in order to describe
 metamaterials with larger lattice constants.

Let us point out that the first term on the right-hand side of
Eq.~\eqref{eq:H_pl-ph} describes, to second-order in perturbation theory, the
exchange of virtual photons among the nanoparticles of the lattice
\cite{craig}. Such a term therefore incorporates the retardation effects in the
dipolar coupling between the LSPs.

\section{Results and discussion}
\label{sec:results}

\subsection{General solution}
\label{sec:PP}
The full Hamiltonian \eqref{eq:H_full}, representing collective plasmons strongly coupled to photons,
can be diagonalized by introducing the bosonic operator
\begin{align}
\label{eq:bogo}
\eta_{\bsl{q}}^{\hat\tau_{\bsl{q}}} =\;& \sum_{\hat\sigma} \left(
u_{\bsl{q}}^{\hat\tau_{\bsl{q}},\hat\sigma} b_{\bsl{q}}^{\hat\sigma}
+v_{\bsl{q}}^{\hat\tau_{\bsl{q}},\hat\sigma} b_{-\bsl{q}}^{\hat\sigma\dagger}
\right) \nonumber\\ &+\sum_{\hat\lambda_\bsl{q}}\left(
m_{\bsl{q}}^{\hat\tau_{\bsl{q}},\hat\lambda_{\bsl{q}}}
c_{\bsl{q}}^{\hat\lambda_{\bsl{q}}}
+n_{\bsl{q}}^{\hat\tau_{\bsl{q}},\hat\lambda_{\bsl{q}}}
c_{-\bsl{q}}^{\hat\lambda_{\bsl{q}}\dagger} \right),
\end{align}
which annihilates a plasmon polariton with wavevector $\bsl{q}$ and
polarization $\hat\tau_{\bsl{q}}$, the latter being generally not aligned with
the $\hat\sigma$-axis.  Imposing that the operator in Eq.~\eqref{eq:bogo} and
its adjoint diagonalize the Hamiltonian \eqref{eq:H_full} as
\begin{equation}
H=\sum_{\bsl{q},\hat\tau_{\bsl{q}}}\hbar\omega_{\mathrm{pp},\bsl{q}}^{\hat\tau_{\bsl{q}}}
\eta_{\bsl{q}}^{\hat\tau_{\bsl{q}}\dagger} \eta_{\bsl{q}}^{\hat\tau_{\bsl{q}}} ,
\end{equation}
the Heisenberg equation of motion $[ \eta_{\bsl{q}}^{\hat\tau_{\bsl{q}}},H]
=
\hbar\omega_{\mathrm{pp},\bsl{q}}^{\hat\tau_{\bsl{q}}}\eta_{\bsl{q}}^{\hat\tau_{\bsl{q}}}$
leads to the $10\times 10$ eigensystem
\begin{widetext}
\begin{equation}
\label{eq:eigensystem}
      \begin{pmatrix}
      \omega_0 \mathbbm{1}_{3}+2\Omega F_\bsl{q} & -2\Omega F_\bsl{q} &
      -\mathrm{i}\omega_0\xi_\bsl{q}P_\bsl{q} &
      \mathrm{i}\omega_{0}\xi_\bsl{q}P_\bsl{q} \\[.1cm] 2\Omega F_\bsl{q} & -(
      \omega_0 \mathbbm{1}_{3}+2\Omega F_\bsl{q})&
      \mathrm{i}\omega_0\xi_\bsl{q}P_\bsl{q} &
      -\mathrm{i}\omega_0\xi_\bsl{q}P_\bsl{q} \\[.1cm]
      \mathrm{i}\omega_0\xi_\bsl{q}P_\bsl{q}^\top &
      \mathrm{i}\omega_0\xi_\bsl{q}P_\bsl{q}^\top & (\omega_{\mathrm{ph},
        \bsl{q}}+2\omega_0\xi_\bsl{q}^2)\mathbbm{1}_{2} &
      -2\omega_0\xi_\bsl{q}^2\mathbbm{1}_{2} \\[.1cm]
      \mathrm{i}\omega_0\xi_\bsl{q}P_\bsl{q}^\top &
      \mathrm{i}\omega_0\xi_\bsl{q}P_\bsl{q}^\top &
      2\omega_0\xi_\bsl{q}^2\mathbbm{1}_{2} & -(\omega_{\mathrm{ph},
        \bsl{q}}+2\omega_0\xi_\bsl{q}^2)\mathbbm{1}_{2}\\
      \end{pmatrix}
          \begin{pmatrix}
              \bsl{u}_{\bsl{q}}^{\hat\tau_{\bsl{q}}}\\
              \bsl{v}_{\bsl{q}}^{\hat\tau_{\bsl{q}}}\\
              \bsl{m}_{\bsl{q}}^{\hat\tau_{\bsl{q}}}\\
              \bsl{n}_{\bsl{q}}^{\hat\tau_{\bsl{q}}}\\
            \end{pmatrix}=
            \omega_{\mathrm{pp},\bsl{q}}^{\hat\tau_{\bsl{q}}}
            \begin{pmatrix}
              \bsl{u}_{\bsl{q}}^{\hat\tau_{\bsl{q}}}\\
              \bsl{v}_{\bsl{q}}^{\hat\tau_{\bsl{q}}}\\
              \bsl{m}_{\bsl{q}}^{\hat\tau_{\bsl{q}}}\\
              \bsl{n}_{\bsl{q}}^{\hat\tau_{\bsl{q}}}\\
            \end{pmatrix},
\end{equation}
\end{widetext}
where the vectors $\bsl{u}_{\bsl{q}}^{\hat\tau_{\bsl{q}}}$,
$\bsl{v}_{\bsl{q}}^{\hat\tau_{\bsl{q}}}$,
$\bsl{m}_{\bsl{q}}^{\hat\tau_{\bsl{q}}}$, and
$\bsl{n}_{\bsl{q}}^{\hat\tau_{\bsl{q}}}$ consist of
$u^{\hat\tau_{\bsl{q}},\hat\sigma}_\bsl{q}$,
$v_\bsl{q}^{\hat\tau_{\bsl{q}},\hat\sigma}$,
$m^{\hat\tau_{\bsl{q}},\hat\lambda_{\bsl{q}}}_\bsl{q}$, and
$n_\bsl{q}^{\hat\tau_{\bsl{q}},\hat\lambda_{\bsl{q}}}$, respectively, as
defined in Eq.\ \eqref{eq:bogo}.  In Eq.~\eqref{eq:eigensystem},
$\mathbbm{1}_{n}$ stands for the $n\times n$ identity matrix, the $3\times3$
symmetric matrix $F_{\bsl{q}}$ is defined by its elements
$f_{\bsl{q}}^{\hat\sigma,\hat\sigma'}$ as given in Eq.~\eqref{fq_mat}, while
the $3\times2$ matrix $P_\bsl{q}$ is introduced as
\begin{equation}
P_\bsl{q}=
 \begin{pmatrix}
\hat x\cdot\hat\lambda_{1, \bsl{q}}&\hspace{.1cm}\hat x\cdot\hat\lambda_{2, \bsl{q}}\\[.1cm]
\hat y\cdot\hat\lambda_{1, \bsl{q}}&\hspace{.1cm} \hat y\cdot\hat\lambda_{2, \bsl{q}}\\[.1cm]
\hat z\cdot\hat\lambda_{1, \bsl{q}}&\hspace{.1cm} \hat z\cdot\hat\lambda_{2, \bsl{q}}
\end{pmatrix},
\end{equation}
and $P_\bsl{q}^\top$ represents its transpose. Here, the two photon
polarizations can be parameterized, e.g., as $\hat\lambda_{1, \bsl{q}}=\hat
z\times\hat q/|\hat z\times\hat q|$ and $\hat\lambda_{2, \bsl{q}}=\hat
q\times\hat\lambda_{1, \bsl{q}}/|\hat q\times\hat\lambda_{1, \bsl{q}}|$ for
$\hat q\nparallel\hat z$, while for $\hat q=\hat z$, we choose $\hat
\lambda_{1, \bsl{q}}=\hat x$ and $\hat \lambda_{2, \bsl{q}}=\hat y$.

We note that the plasmon-polariton eigenfrequencies
$\omega_{\mathrm{pp},\bsl{q}}^{\hat\tau_{\bsl{q}}}$ arising from the
eigensystem \eqref{eq:eigensystem} occur in pairs of positive and negative
eigenvalues. Below, we will focus on the physically relevant, positive
solutions.

If not stated otherwise, we will use an interparticle distance
$a=3r_{\mathrm{np}}$, a cutoff radius $\rho_\mathrm{c}=150 a$, and
$\alpha=10$. We have checked that the latter choices provide
numerically-converged results for the collective plasmon and plasmon-polariton
dispersions, presented in the next subsections.

\subsection{Collective plasmons}
\label{sec:CP}

Before considering the fully coupled system, represented by the Hamiltonian
\eqref{eq:H_full}, it is instructive to analyze in detail the purely plasmonic
problem described by $H_\mathrm{pl}$ in Eq.~\eqref{eq:H_pl}. We will therefore
set the light-matter coupling to zero in this subsection. In this way, plasmon
properties are computed in the quasistatic limit, neglecting all retardation
effects.

\begin{figure}[tb]
  \begin{center}
       \includegraphics[keepaspectratio=true, width=\columnwidth]{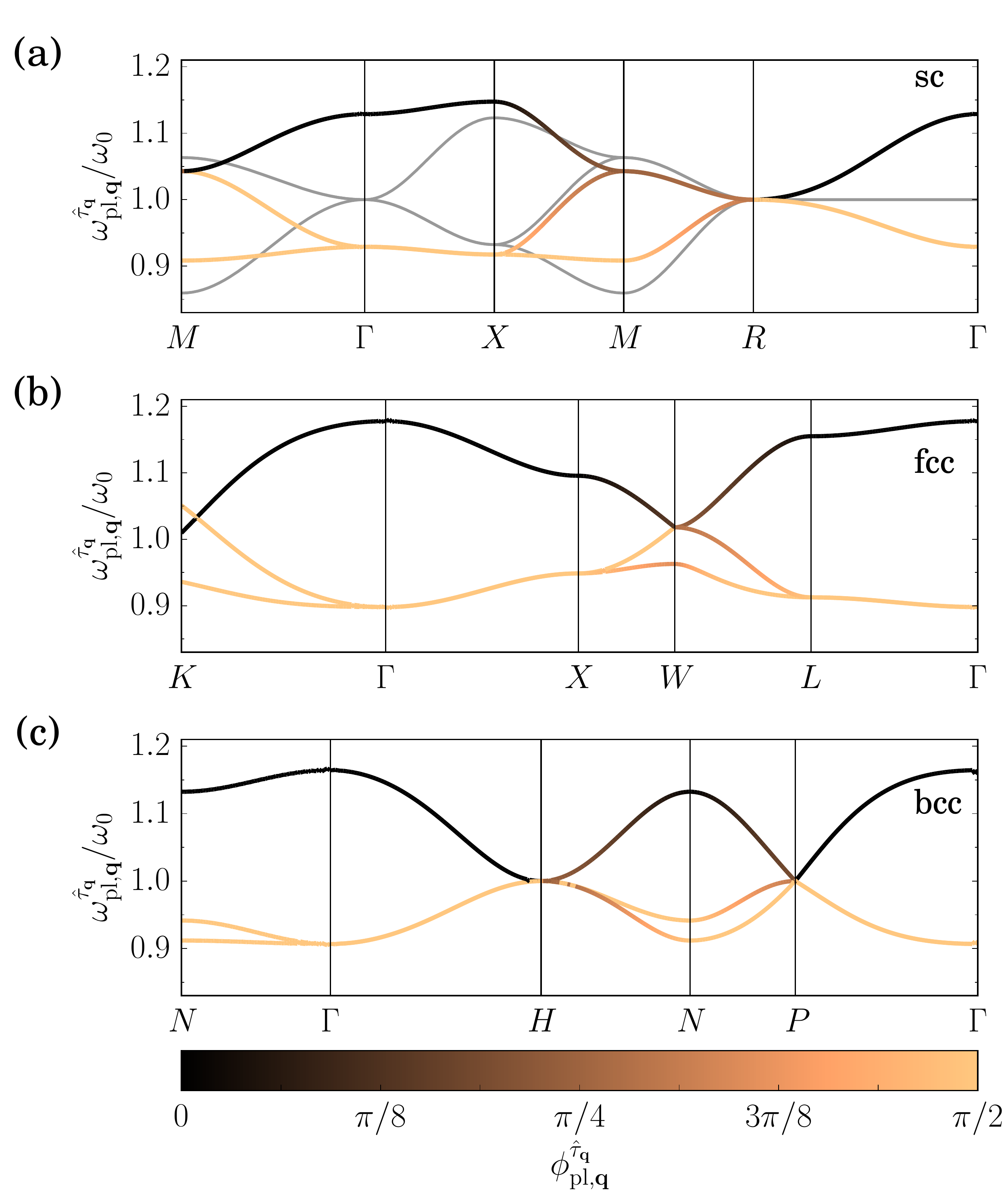}
  \end{center}
\caption{Collective plasmon dispersion
  $\omega_{\mathrm{pl},\bsl{q}}^{\hat\tau_{\bsl{q}}}$ in units of the LSP
  frequency $\omega_0$ along the paths shown in red in
  Figs.~\ref{fig:latbzs}(d)-(f) for the (a) sc, (b) fcc, and (c) bcc
  lattices. The color code corresponds to the collective plasmon-polarization
  angle $\phi_{\mathrm{pl},\bsl{q}}^{\hat\tau_{\bsl{q}}}$, which equals $0$
  ($\pi/2$) for purely longitudinal (transverse) plasmons.  In the figure we
  use $a=3r_{\mathrm{np}}$, $\rho_{\mathrm{c}}=150a$ and $\alpha=10$ for the
  colored thick lines, while in panel (a) we choose $\rho_{\mathrm{c}}=a$ and
  $\alpha=0$ for the gray thin lines, corresponding to nearest-neighbor
  interactions only [cf.~Eq.~\eqref{eq:omega_pl_sc_nn}]. In all cases the
  dielectric constants are set to $\epsilon_{\mathrm{d}}= 5.6$ and
  $\epsilon_{\mathrm{m}}= 4$.}
\label{pl-dis}
\end{figure}

Setting $\xi_\bsl{q}=0$, the matrix defined in Eq.~\eqref{eq:eigensystem}
becomes block-diagonal. On the one hand, the lower $4\times4$ block is
diagonal and corresponds to the two degenerate photon branches with dispersion
$\omega_{\mathrm{ph},\bsl{q}}$ for the two positive eigenvalues. The three
positive eigenvalues of the upper $6\times6$ block, on the other hand, yield
the collective plasmon dispersion
$\omega_{\mathrm{pl},\bsl{q}}^{\hat\tau_{\bsl{q}} }$, which is represented in
Fig.~\ref{pl-dis} as a function of wavevector $\bsl{q}$ along the red paths
given in Figs.~\ref{fig:latbzs}(d)-(f) for the sc [Fig.~\ref{pl-dis}(a)], fcc
[Fig.~\ref{pl-dis}(b)], and bcc [Fig.~\ref{pl-dis}(c)] lattices.  In the
figure we use $\epsilon_{\mathrm{d}}= 5.6$, as determined for silver films
\cite{Blaber2009,Yang2015}, and $\epsilon_{\mathrm{m}}= 4$, mimicking an
embedding medium made of glass or polymer.  In Fig.~\ref{pl-dis} we also show
the collective plasmon-polarization angle
$\phi_{\mathrm{pl},\bsl{q}}^{\hat\tau_{\bsl{q}}}=\arccos{(|\hat{\tau}_{\bsl{q}}\cdot\hat{q}|)}$,
where we choose
$\hat{\tau}_{\bsl{q}}=\hat{{u}}_{\bsl{q}}^{\hat\tau_{\bsl{q}}}$. Notice that
the alternative choice
$\hat{\tau}_{\bsl{q}}=\hat{{v}}_{\bsl{q}}^{\hat\tau_{\bsl{q}}}$ leads to the
same polarization angle, as the vectors
$\bsl{u}_{\bsl{q}}^{\hat\tau_{\bsl{q}}}$ and
$\bsl{v}_{\bsl{q}}^{\hat\tau_{\bsl{q}}}$ are proportional for a given
wavevector $\bsl{q}$. With the above definition of
$\phi_{\mathrm{pl},\bsl{q}}^{\hat\tau_{\bsl{q}}}$, longitudinal collective
plasmons, which do not couple to light, have a polarization angle
$\phi_{\mathrm{pl},\bsl{q}}^{\hat\tau_{\bsl{q}}}=0$ (black lines in
Fig.~\ref{pl-dis}), while purely transverse modes have a corresponding polarization
$\phi_{\mathrm{pl},\bsl{q}}^{\hat\tau_{\bsl{q}}}=\pi/2$ (yellow lines in
Fig.~\ref{pl-dis}).

Our results in Fig.~\ref{pl-dis} indicate that there are two purely transverse
collective plasmons and one purely longitudinal one along the high-symmetry
axes in the first Brillouin zone [i.e., axes with 2- to 4-fold rotational
  symmetry, see Figs.~\ref{fig:latbzs}(d)-(f)]. For less symmetric axes the
collective modes can be of a mixed type [see, e.g., the $XM$ and $MR$ lines in
  Fig.~\ref{pl-dis}(a)]. Moreover, along 3- and 4-fold symmetry axes, the two
transverse modes are degenerate [see, e.g., the $\Gamma R$ and $\Gamma X$
  lines in Fig.~\ref{pl-dis}(a)]. This is a manifestation of Neumann's
principle~\cite{nye}: For the collective plasmon dispersion this enforces the
degeneracy of the transverse modes for the 3- and 4-fold symmetry lines. The
latter degeneracy is lifted for wavevector directions with lower symmetry.
We also note that one would expect the longitudinal and transverse plasmon
  modes to be degenerate at the $\Gamma$ point since the latter has the full
  point-group symmetry of the lattice.  As we will see later, there is a
  radiative correction from the light-matter interaction Hamiltonian
  \eqref{eq:H_pl-ph} that enforces this degeneracy in the polariton spectrum.

Before we move on to the discussion of the fully coupled system, a comment is
in order about the importance of the dipole-dipole interaction beyond nearest
neighbors for the collective plasmon dispersion.  In Fig.~\ref{pl-dis}(a) we
represent by thin gray lines the plasmon dispersion of the sc lattice,
including nearest-neighbor interactions only. (Note that we do not correct for
the Wilbraham-Gibbs phenomenon around the $\Gamma$ point in this case,
i.e., we use $\alpha=0$.)  Under these conditions, the matrix $F_\bsl{q}$ is
diagonal, and its elements read
\begin{equation}
  f_\bsl{q}^{\hat\sigma,\hat\sigma'}=\delta_{\hat\sigma\hat\sigma'}
  \sum_{\hat\sigma''=\hat x, \hat y, \hat z}(1-3\delta_{\hat\sigma\hat\sigma''})\cos{(a\hat\sigma''\cdot\bsl{q})}.
\end{equation}
The plasmonic Hamiltonian \eqref{eq:H_pl} is therefore separable into $\hat
x$, $\hat y$ and $\hat z$ directions and can be diagonalized analytically,
yielding
\begin{equation}
  \label{eq:omega_pl_sc_nn}
  \omega_{\mathrm{pl},\bsl{q}}^{\hat\sigma}=\omega_0
  \sqrt{1+4\frac{\Omega}{\omega_0} f_\bsl{q}^{\hat\sigma,\hat\sigma}}.
\end{equation}
This result and the corresponding coefficients of the Bogoliubov
transformation \eqref{eq:bogo}, which we do not report explicitly here,
coincide with those found in Ref.~\cite{Weick} for LSP polarizations along
$\hat x$, $\hat y$ or $\hat z$ and
$\epsilon_{\mathrm{m}}=\epsilon_{\mathrm{d}}=1$.  As can be seen in
Fig.~\ref{pl-dis}(a), including the dipole-dipole interactions beyond nearest
neighbors can have a qualitative effect on the collective plasmon dispersion,
most noticeably around the $\Gamma$ point.  They further lift the degeneracy
between plasmon branches, e.g., along the $\Gamma M$ and $\Gamma R$
directions.  In other regions of the first Brillouin zone the difference
between the full dispersion and those from nearest neighbors only is less
significant.

\subsection{Plasmon polaritons}
\label{sec:full}

\begin{figure*}[tb]
  \begin{center}
    \includegraphics[keepaspectratio=true,
      width=\linewidth]{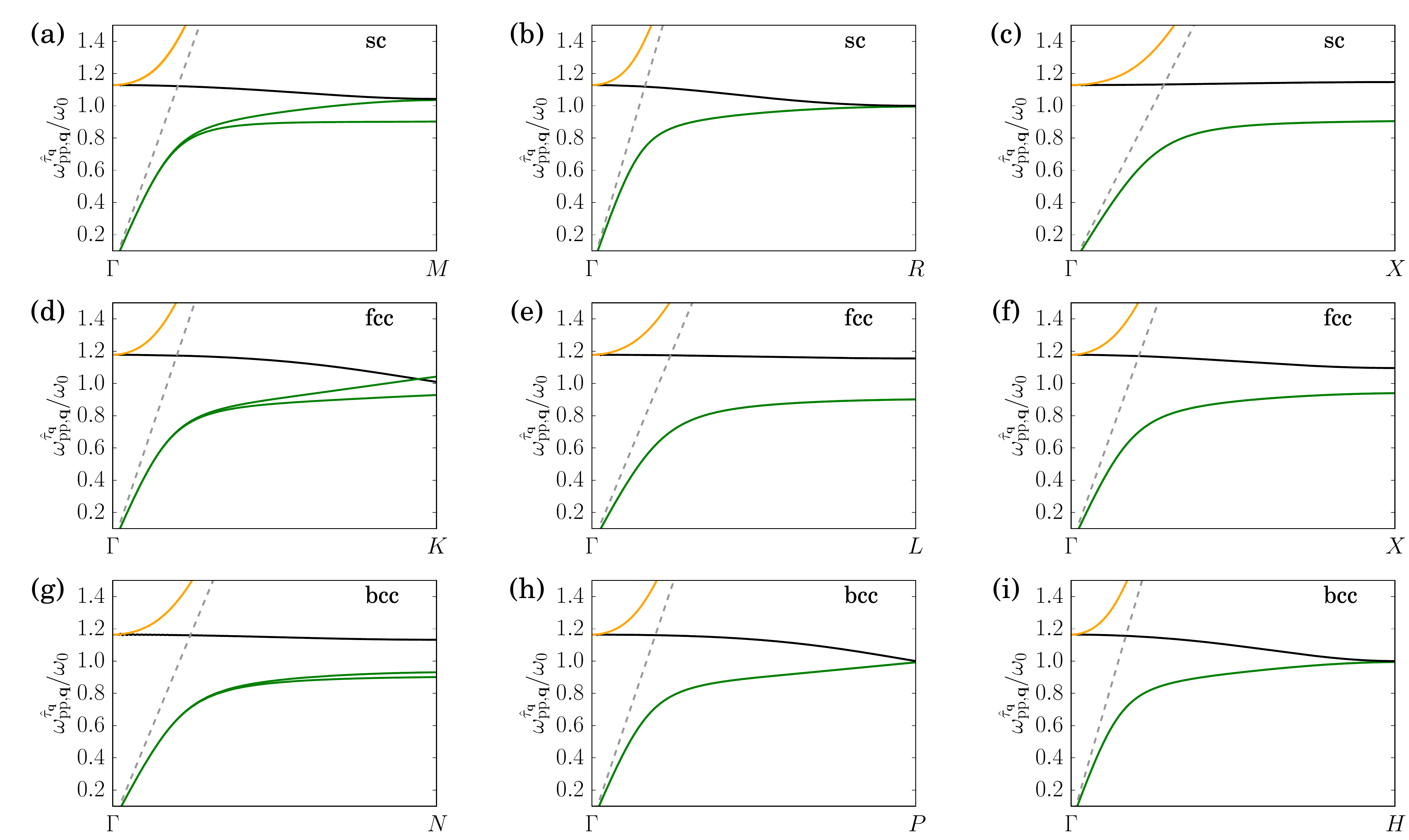}
  \end{center}
  \caption{Solid lines: plasmon-polariton dispersion $\omega_{\mathrm{pp},
      \bsl{q}}^{\hat\tau_\bsl{q}}$ in units of the LSP frequency $\omega_0$
    for the (a)-(c) sc, (d)-(f) fcc, and (g)-(i) bcc lattices along the
    (a),(d),(g) 2-fold, (b),(e),(h) 3-fold, and (c),(f),(i) 4-fold symmetry
    axes, shown in Fig.~\ref{fig:latbzs}. Dashed gray lines: free photon
    dispersion $\omega_{\mathrm{ph},\bsl{q}}$.  The parameters used in the
    figure are $a=3r_{\mathrm{np}}$, $\rho_\mathrm{c}=150a$, $\alpha=10$,
    $\omega_0r_{\mathrm{np}}/c=0.237$, $\epsilon_{\mathrm{d}}= 5.6$ and
    $\epsilon_{\mathrm{m}}= 4$.}
  \label{pp-dis}
\end{figure*}

We now consider the fully coupled system, represented by the eigensystem
\eqref{eq:eigensystem}, and numerically solve for its five positive
eigenvalues. These eigenvalues yield the plasmon-polariton spectrum
$\omega_{\mathrm{pp},\bsl{q}}^{\hat\tau_\bsl{q}}$, which is shown by solid
lines in Fig.~\ref{pp-dis} for the sc [Figs.~\ref{pp-dis}(a)-(c)], fcc
[Figs.~\ref{pp-dis}(d)-(f)], and bcc lattices [Figs.~\ref{pp-dis}(g)-(i)]
along 2-fold [Figs.~\ref{pp-dis}(a),(d),(g)], 3-fold
[Figs.~\ref{pp-dis}(b),(e),(h)], and 4-fold symmetry axes
[Figs.~\ref{pp-dis}(c),(f),(i)], cf.\ Figs.~\ref{fig:latbzs}(d)-(f).  Along
the high symmetry axes of the first Brillouin zone, the five modes split up
into four polaritonic branches (colored solid lines) and one purely
longitudinal collective plasmon, which does not couple to transverse photons
(black lines). The four polaritonic modes result from the coupling of
transverse collective plasmons (see Fig.~\ref{pl-dis}) to photons, whose
dispersion relation is shown by dashed lines in Fig.~\ref{pp-dis}. According
to the construction of our effective model and the nature of the
Coulomb gauge, retardation effects are taken into account for all
plasmon-polariton branches, where photons and plasmons interact via
Eq.~(\ref{eq:H_pl-ph}).

As can be inferred from Fig.~\ref{pp-dis}, there are two high-energy
polaritonic branches (orange solid lines) and two low-energy ones (green solid
lines). The two high-energy branches are nearly degenerate.  The low-energy
polaritonic branches, shown by green solid lines in Fig.~\ref{pp-dis}, have
the same twofold degeneracy along 3-fold and 4-fold symmetry axes as the
collective plasmon dispersion (compare with Fig.~\ref{pl-dis}) and the
light-matter interaction does not lift this degeneracy.  As mentioned
  previously, there is a radiative correction to the transverse plasmonic
  modes at the $\Gamma$ point, which is equal to the longitudinal-transverse
  splitting observed in the plasmonic spectrum (Fig.\ \ref{pl-dis}). As a
  result, one observes that the longitudinal and transverse high-energy
  polaritonic branches are degenerate at the $\Gamma$ point
  (Fig.\ \ref{pp-dis}), as required by symmetry.

For wavevectors close to the edge of the first Brillouin zone the high-energy
polaritonic branches (orange solid lines in Fig.~\ref{pp-dis}) asymptotically
approach the light cone, while the low-energy ones (green solid lines in the
figure) tend to the collective plasmon dispersion. For
$\mathbf{q}\rightarrow0$ (i.e., close to the $\Gamma$ point), the states
corresponding to the low-energy branches are mostly photon-like, with a
renormalized group velocity, which is smaller than
$c/\sqrt{\epsilon_{\mathrm{m}}}$, indicating an effective index of refraction
larger than $\sqrt{\epsilon_{\mathrm{m}}}$. However, the high-energy
  branches do not tend to the values displayed in Fig.\ \ref{pl-dis} at the
  $\Gamma$ point due to the strong coupling between collective plasmons and
  photons [cf.~Eq.~\eqref{eq:H_pl-ph}]. This results in a splitting between
  the low- and high-energy polaritonic branches. We
define this polaritonic splitting $\Delta_{\hat q}$ as the frequency
difference between the minimum of the high-energy polaritonic branches and the
maximum of the lower branches over all wavevectors $\bsl{q}$ in the first
Brillouin zone along a fixed direction $\hat q$ from the $\Gamma$ point.

As can be seen in Fig.~\ref{pp-dis} for $\epsilon_{\mathrm{d}}=5.6$ and
$\epsilon_{\mathrm{m}}=4$, the polaritonic splitting reaches values of the
order of $\unit[25]{\%}$ of the LSP resonance frequency $\omega_0$. For noble-metal
nanoparticles the latter typically lies in the visible to ultraviolet range
($\omega_0\simeq2$--$\unit[4]{eV/\hbar}$), resulting in a splitting of about
$\Delta_{\hat q}\simeq0.5$--$\unit[1.0]{eV/\hbar}$.  The splitting in the
polaritonic dispersion has important experimental consequences for the optical
properties of the metamaterial. Indeed, along a certain direction $\hat q$ in
the Brillouin zone, no plasmon polariton can propagate for frequencies within
the bandgap, so that the reflectivity of the metacrystal should be perfect. We would like to emphasize that the physical origin of these band gaps is entirely different from those emerging in conventional photonic crystals which are the result of Bragg scattering \cite{braggscat}. In fact, we neglect Umklapp processes and therefore the band gaps emerge as a result of polaritonic hybridization between Mie resonances and photons.

Interestingly, the polaritonic splitting depends on the polarization for the
two-fold symmetry axes of the three cubic lattices [see
  Figs.~\ref{pp-dis}(a),(d),(g)].  This birefringence is directly related to
the polarization dependence of the collective plasmon dispersion, the latter
being due to the anisotropic nature of the dipole-dipole interaction between
the nanoparticles composing the metamaterial. The modulation of the band
splitting can be rather significant for the sc and fcc lattices (around $\unit[12]{\%}$
of $\omega_0$), while for the bcc lattice it is comparatively less (around $\unit[3]{\%}$
of $\omega_0$). In the following, we will refer to the modulation of
$\Delta_{\hat q}$ for different polarizations as $\delta_{\hat q}$.

\begin{figure}[t!]
  \begin{center}
\includegraphics[keepaspectratio=true,
  width=.95\linewidth]{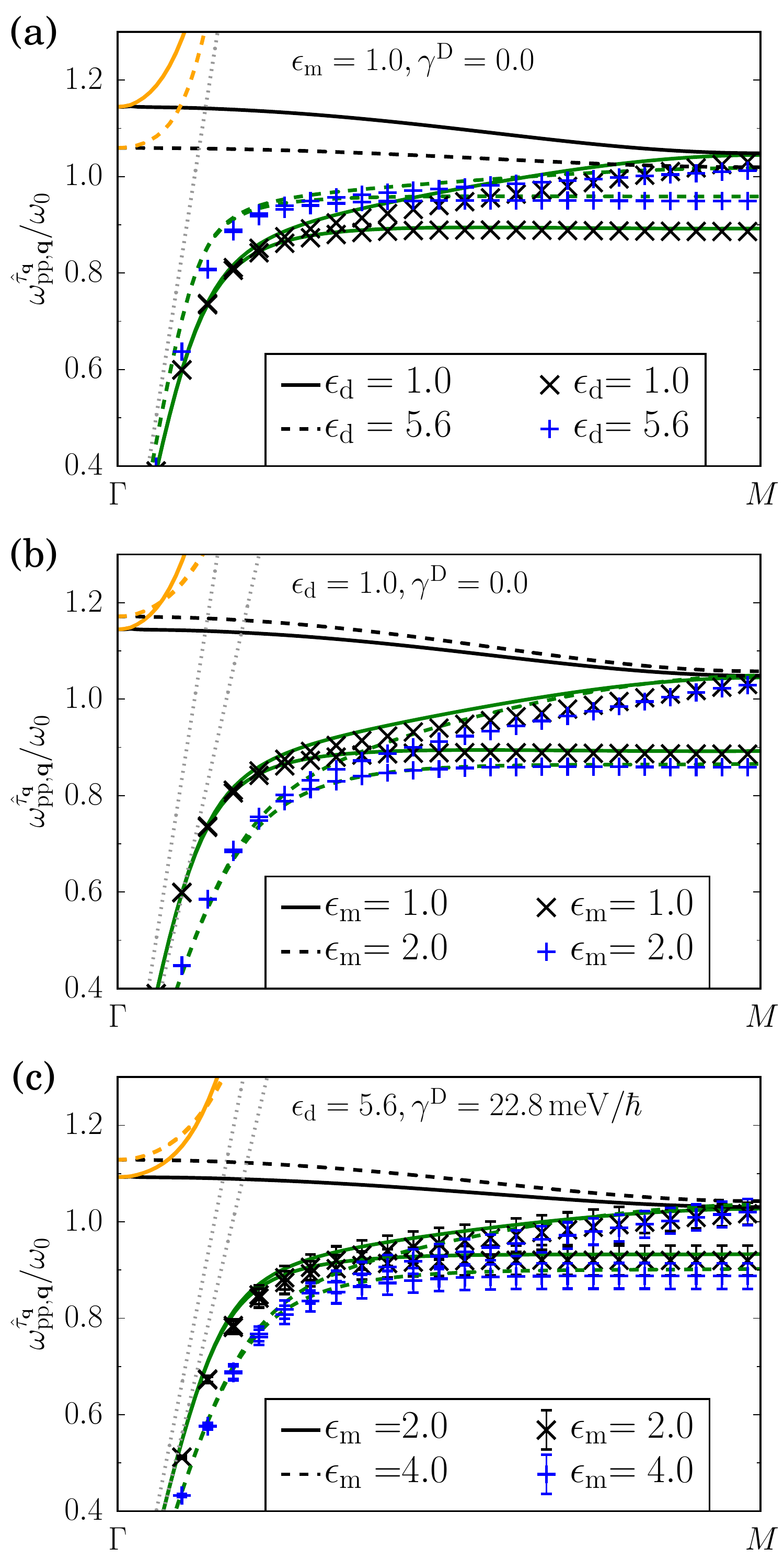}
  \end{center}
\caption{Plasmon-polariton dispersions for various values of the
  dielectric constants $\epsilon_{\mathrm{d}}$ and
  $\epsilon_{\mathrm{m}}$. The lines present the results of our Hamiltonian
  approach [see Eq.~\eqref{eq:eigensystem}] and the symbols those found as
  solutions in FDFD simulations. Solid and dashed lines: plasmon-polariton
  dispersions $\omega_{\mathrm{pp}, \bsl{q}}^{\hat\tau_\bsl{q}}$ in units of
  the LSP frequency $\omega_0$ for the sc lattice along the 2-fold symmetry
  axis (see Fig.~\ref{fig:latbzs}). Dotted gray lines: free photon dispersions
  $\omega_{\mathrm{ph},\bsl{q}}$.  The parameters for the Hamiltonian approach are
   $a=3r_{\mathrm{np}}$, $\alpha=10$, and $\rho_\mathrm{c}=150a$, while
  we choose  $r_{\mathrm{np}}=\unit[10]{nm}$ and $a=3r_{\mathrm{np}}$
  in the FDFD calculations. In panels (a) and (b) the LSP frequency
  $\omega_0r_{\mathrm{np}}/c=0.177$ [see Eq.~\eqref{eq:omegaLSP}] is kept
  constant, while in panel (c) $\omega_{\mathrm{p}}=\unit[9.6]{eV/\hbar}$ is
  constant. All other parameters are indicated in the respective panels.
  In the FDFD calculations presented in panel (c), we choose a finite Drude damping
  $\gamma^{\mathrm{D}}$~\cite{Blaber2009}, and plot the calculated imaginary
  parts of the eigenfrequencies, i.e., damping rates, as error bars.}
\label{fig:QM+Comsol}
\end{figure}

Let us now discuss the dependence of the plasmon-polariton dispersions on the
dielectric constants $\epsilon_{\mathrm{d}}$ and $\epsilon_{\mathrm{m}}$ for
the two-fold symmetry axes, as shown in Fig.~\ref{fig:QM+Comsol}. To simplify
the discussion, we keep the Mie frequency $\omega_0$ constant in
Figs.~\ref{fig:QM+Comsol}(a) and \ref{fig:QM+Comsol}(b) by adjusting
$\omega_{\mathrm{p}}$, while the Mie frequency is varied in
Fig.~\ref{fig:QM+Comsol}(c). As indicated in Fig.~\ref{fig:QM+Comsol}(a), an
increase in the screening of the core electrons decreases the polaritonic
splitting and leads to a corresponding flattening of the longitudinal plasmon
branch. The smaller splitting can be understood by noting that the coupling
constant $\Omega\propto 1/(2+\epsilon_{\mathrm{d}}/\epsilon_{\mathrm{m}})$ in
the plasmonic part [see Eq.~\eqref{eq:Omega}] decreases with increasing
$\epsilon_{\mathrm{d}}$. The dependence of the polaritonic dispersion on the
dielectric constant of the surrounding medium $\epsilon_{\mathrm{m}}$ is more
complex, as displayed in Fig.~\ref{fig:QM+Comsol}(b). An increasing
$\epsilon_{\mathrm{m}}$ reduces the effective speed of light in the
medium. Hence, this reduces the slope of the low-energy polaritonic branches
around the $\Gamma$-point, while the slope of the high-energy polaritonic
branches is modified away from the $\Gamma$ point. Furthermore, with
increasing $\epsilon_{\mathrm{m}}$ a larger polaritonic splitting
$\Delta_{\hat q}$ as well as increased modulation $\delta_{\hat q}$ between
the low-energy polaritonic branches of different polarization is observed. We
attribute this to two factors. Most importantly the coupling constant $\Omega$
increases with increasing $\epsilon_{\mathrm{m}}$, and thus the related band
splittings get larger. An increasing $\epsilon_{\mathrm{m}}$ also enhances the
plasmon-photon coupling as $\xi_{\bsl{q}}\propto
\epsilon_{\mathrm{m}}^{1/4}/(2+\epsilon_{\mathrm{d}}/\epsilon_{\mathrm{m}})^{1/2}$
        [see Eq.~\eqref{eq:H_pl-ph}], but the effect of $\xi_{\bsl{q}}$ on the
        polaritonic dispersion is not easily quantified. In
        Fig.~\ref{fig:QM+Comsol}(c) the plasma frequency $\omega_{\mathrm{p}}$
        is fixed to the value of silver films~\cite{Blaber2009,Yang2015},
        while the dielectric constant $\epsilon_{\mathrm{m}}$ of the medium is
        varied. In this case, we observe similar effects as in
        Fig.~\ref{fig:QM+Comsol}(b).
        
We note that for certain high-symmetry axes it is possible to derive
  analytic expressions for the components of the dielectric tensor of the
metamaterial, as we show in the Appendix. Their dependencies on the wavevector
and frequency indicate a nonlocal behavior of the metamaterial in space and
time.

The experimental observability of the band splittings $\Delta_{\hat q}$ and of
their polarization-dependent modulation $\delta_{\hat q}$, discussed above,
may be hindered by damping mechanisms, leading to the decay of the plasmon
polaritons. The latter are mostly subject to two sources of damping: Ohmic
(absorption) losses with decay rate $\gamma^\mathrm{D}$ inherent to any type
of metallic nanostructure [see Eq.~(\ref{eq:epsDrude})], and Landau damping
with decay rate $\gamma^{\mathrm{L}}$, i.e., the decay of the plasmon
excitation into electron-hole pairs \cite{kreibig, kawab66_JPSJ}. Note that
radiation damping is irrelevant for the infinite metacrystals considered here
since there is no photonic continuum into which the plasmons can decay. Ohmic losses were experimentally estimated to
be of the order of $\gamma^\mathrm{D}\simeq\unit[24]{meV/\hbar}$ for bulk
silver~\cite{Blaber2009}. Moreover, it
has been shown that Landau damping only weakly depends on the dipole-dipole
interaction~\cite{brand15_PRB, brand16_PRB, Downing2017}, so that we estimate
it with the Landau damping of a single nanoparticle. This yields
$\gamma^\mathrm{L}=3 v_{\mathrm{F}}g/4r_{\mathrm{np}}$, where $v_{\mathrm{F}}$
is the Fermi velocity and $g$ is a numerical factor of the order of $1$
\cite{kreibig, kawab66_JPSJ, yanno92_AP, weick05_PRB}. For Ag nanoparticles,
we obtain
$\hbar\gamma^\mathrm{L}\simeq\unit[690]{meV}/r_{\mathrm{np}}\mathrm{[nm]}$. For
the nanoparticle radii that we consider (typically of the order of
\unit[10]{nm}), the total linewidth of the plasmon-polariton bandstructure is
therefore of the order of
$\gamma^\mathrm{D}+\gamma^\mathrm{L}\simeq\unit[100]{meV}/\hbar$. For this
reason the splittings in the plasmon-polariton dispersion $\Delta_{\hat q}$, as
well as their polarization dependence $\delta_{\hat q}$ for certain directions
in the first Brillouin zone, should be experimentally accessible.

\section{Comparison to classical electrodynamics simulations}
\label{sec:compare}

To validate the predictions of our Hamiltonian approach presented in the
preceding section, we compare them here to calculations based on classical
electrodynamics. FDFD simulations are carried out with the electromagnetic
wave module of the \textsc{COMSOL Multiphysics} package with the
eigenfrequency solver. We numerically search for solutions to the
eigenequation
\begin{equation}
  \bs{\nabla}\times\left[\bs{\nabla} \times
  \bsl{E}(\bsl{r},\omega)\right]-\left(\frac{\omega}{c}\right)^2\epsilon_{\mathrm{r}}(\bsl{r},\omega)\bsl{E}(\bsl{r},\omega)
  =0,
\end{equation}
where $\bsl{E}(\bsl{r},\omega)$ corresponds to the electric field at position
$\bsl{r}$ and frequency $\omega$, and where
$\epsilon_{\mathrm{r}}(\bsl{r},\omega)$ characterizes the dielectric
properties of the metamaterial.  We consider an infinite, sc lattice with
  a lattice constant of $\unit[30]{nm}$, which allows us to simplify the
  numerical calculations by applying Floquet periodicity on the faces of a
  unit cell for the electric and magnetic fields. We choose nanoparticles of
  radius $\unit[10]{nm}$ and model them using the Drude dielectric function of
  Eq.~\eqref{eq:epsDrude}, while in the embedding medium
  $\epsilon_{\mathrm{r}}(\bsl{r},\omega)=\epsilon_{\mathrm{m}}$. Note that
  since we use the eigenvalue solver in \textsc{COMSOL}, we do not insert a
  driving source into the system.  The meshes on three surfaces of the cubic
  cell are of a free triangular type. They are copied to the opposite side to
  be compatible with the Floquet periodicity.  The cubic cell is filled with
  an automatically-generated tetrahedral mesh, and the parameters utilized for
  generating the triangular and tetrahedral meshes are listed in Table
  \ref{tab}.

\begin{table}
\caption{\label{tab}  Parameters of the triangular and tetrahedral meshes used in the \textsc{COMSOL} simulations.}
\begin{ruledtabular}
\begin{tabular}{lc}
Maximum element size & $\unit[2.4]{nm}$\\
Minimum element size & $\unit[0.3]{nm}$\\
Maximum element growth rate & 1.45\\
Curvature factor & $0.5$\\
Resolution of narrow regions & $0.6$\\
Geometry scaling & $1$\\
Adaptive mesh refinement & not used
\end{tabular}
\end{ruledtabular}
\end{table}

The results of the FDFD calculations for the low-energy polaritonic branches
are summarized with symbols in Fig.~\ref{fig:QM+Comsol}. As for the
Hamiltonian approach, parameters in Figs.~\ref{fig:QM+Comsol}(a) and
\ref{fig:QM+Comsol}(b) are adjusted to give the same Mie frequency
$\omega_0=\unit[3.48]{eV}/\hbar$ for the nanoparticles.  In
Fig.~\ref{fig:QM+Comsol}(c) we keep $\omega_{\mathrm{p}}$ and
$\epsilon_{\mathrm{d}}$ constant, varying $\epsilon_{\mathrm{m}}$ and
exploring the influence of a finite Drude damping $\gamma^{\mathrm{D}}$, which
is not contained in our Hamiltonian-based model. The parameters
$\omega_{\mathrm{p}}$ and $\gamma^{\mathrm{D}}$ are chosen as specified for
silver in Ref.~\cite{Blaber2009}. We find an excellent agreement of the FDFD
simulations with the predictions of our effective model in all cases,
confirming its validity. To avoid repetition, we refrain from discussing in
further detail the results of the FDFD calculations in
Figs.~\ref{fig:QM+Comsol}(a) and \ref{fig:QM+Comsol}(b), but concentrate on
the new aspect due to the inclusion of a finite damping in
Fig.~\ref{fig:QM+Comsol}(c). There, the imaginary part of the
eigenfrequencies, which can be interpreted as the linewidth broadening due to
Ohmic losses, is represented by error bars. We find a general trend of an
increased damping with increasing wavevector. Since the broadenings turn out
to have nearly no influence on the polaritonic dispersion relations, our model
reproduces the dispersions with great accuracy. The small red shift of the
FDFD calculations with respect to our model can be understood by 
the fact that we neglect Umklapp scattering and higher-order multipolar bands, which would push the bands downward in energy. 
Even if a
wavevector-independent broadening $\gamma^{\mathrm{L}}$ due to Landau damping
would be added, which we argued to be actually larger than the broadening due
to Ohmic losses (see the discussion in Sec.~\ref{sec:full}), the
polarization-dependent band gap modulation $\delta_{\hat q}$ should still be
observable.

\begin{figure}[tb]
  \begin{center}
    \includegraphics[keepaspectratio=true, width=\linewidth]{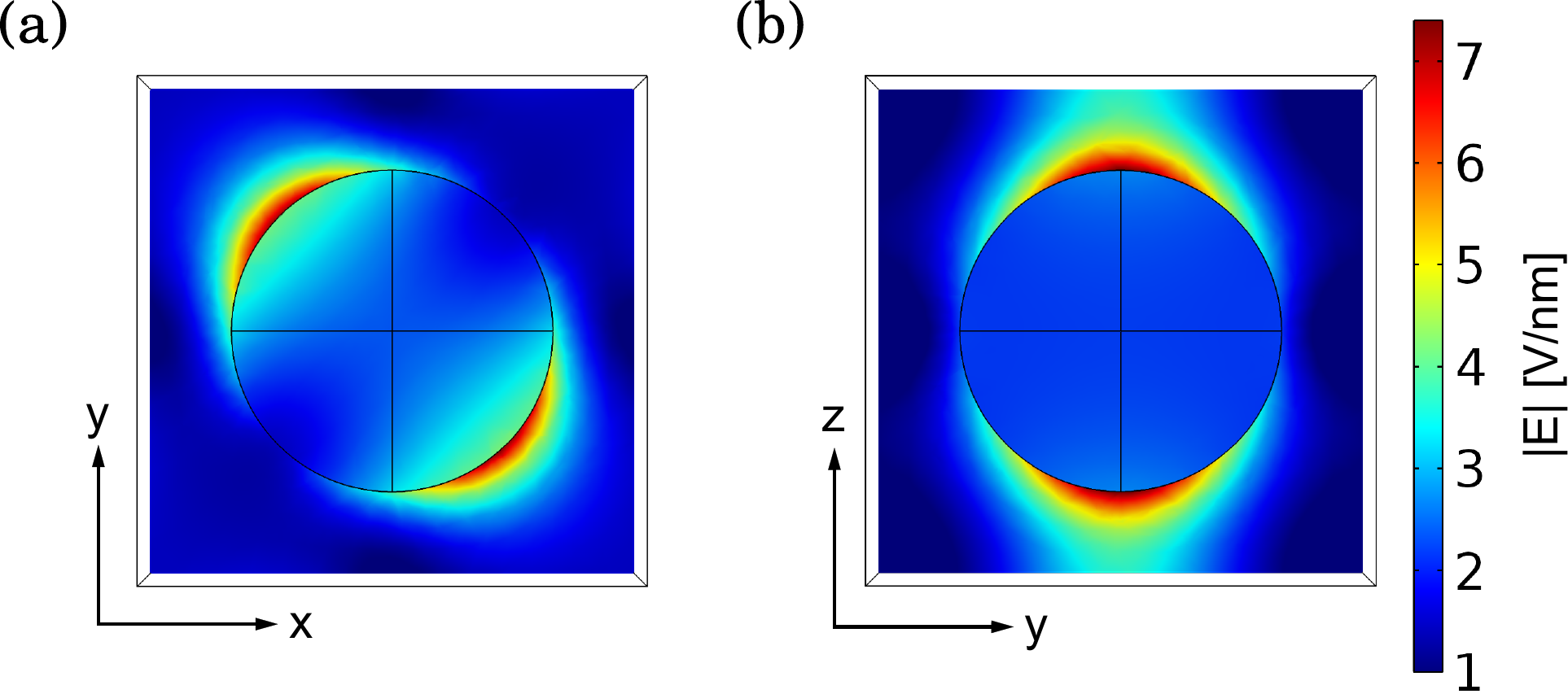}
  \end{center}
  \caption{(a),(b) Density plots for the distribution of the electric field of
    the low-energy transverse polaritonic modes, shown on a plane cutting
    through the center of the primitive cell. These distributions are
    calculated at $\bsl{q}=0.49\pi(\hat x+\hat y)/a$ in the direction of the
    twofold symmetry axis, using the parameters $r_{\mathrm{np}}=\unit[10]{nm}$,
    $a=3r_{\mathrm{np}}$, $\epsilon_{\mathrm{d}}=5.6$,
    $\epsilon_{\mathrm{m}}=1$, $\omega_{\mathrm{p}}=\unit[9.6]{eV/\hbar}$ and
    $\gamma^\mathrm{D}=\unit[22.8]{meV/\hbar}$ \cite{Blaber2009}. The dipolar modes
    exhibit a polarization oriented in the (a) $\hat y - \hat x$ and (b) $\hat
    z$ directions.}
  \label{fig:cmodes}
\end{figure}

With the distribution of the electric field available in the \textsc{COMSOL} package,
we can check the polarization direction that our Hamiltonian approach predicts
for the sc lattice. Along the $\Gamma M$ direction with $\hat q=(\hat x+\hat
y)/\sqrt{2}$ [see Figs.~\ref{fig:latbzs}(d) and \ref{fig:QM+Comsol}] we find
that the lowest-energy transverse plasmon-polariton branch exhibits a
polarization $\hat \tau_{\bsl{q}}$ parallel to the $\hat z$-axis, while the
second lowest-energy one exhibits a polarization $\hat{\tau}_{\bsl{q}}$
parallel to $\hat y - \hat x$. This is indeed confirmed by the FDFD
calculations for all the parameter sets tested in Fig.~\ref{fig:QM+Comsol}. An
example of the field distributions is given in Figs.~\ref{fig:cmodes}(a) and
\ref{fig:cmodes}(b). For different lengths of the reciprocal wavevectors $q$,
these modes change in details like the field distribution in the middle of the
nanoparticle or the calculated field strength, but the polarization directions
and the overall dumbbell shape remain the same.

We would like to highlight some difficulties in obtaining the polariton
dispersions using FDFD calculations. As we are trying to solve a non-linear
equation in 3D, the solver also converges on many unphysical solutions where
the electric field is, e.g., concentrated in a single spot or exhibits a
  random-looking distribution. As a result, one has to manually inspect the
  field profile of the eigenmodes, discarding the artificial solutions and
  retaining only those with a dipolar-like character, such as the ones shown
in Fig.\ \ref{fig:cmodes}. Also, the starting and linearization points for the
eigenfrequency search were varied for the different calculations, and we
checked that they had negligible effect on the real part of the
eigenfrequency. However, we find that the imaginary part is less robust and
changes with distance of the real part of the eigenfrequency from the
linearization point. For this reason, we took care that the linearization
points were located close to the respective eigenfrequencies at each
wavevector.

Furthermore, in the discussion above, we have focused on the low-energy
polariton branches which show the interesting polarization-dependent band
splittings. However, at higher frequencies one can also find, in addition to
the longitudinal branch, many polaritonic branches which arise from
  multipolar plasmon modes.

The excellent agreement of plasmon-polariton dispersions predicted by our
Hamiltonian-based model with those of the FDFD simulations shows that higher
multipolar modes beyond the considered dipolar interactions as well as
intraparticle retardation effects are irrelevant in the studied parameter
regime. In addition, we want to point out that the computational costs of our
Hamiltonian approach are only a fraction of those of the \textsc{COMSOL}
simulations and avoid the cumbersome problems related to the convergence
  to unphysical solutions. Our Hamiltonian-based approach is thus an
efficient way to quantitatively predict the response of metacrystals in the
near-field regime, when meta-atom separations are much smaller than the LSP
resonance wavelength, i.e., $\omega_0 a/c\ll1$.

\section{Conclusions}
\label{sec:ccl}
In this work we have theoretically studied plasmon polaritons in sc, fcc and
bcc lattices of spherical metallic nanoparticles. We have developed a model
based on a quantum-mechanical Hamiltonian, justified for small nanoparticles
(i.e., with a radius between ca.~$1$ and $\unit[20]{nm}$) in the near-field
dipolar regime. The dipole-dipole interaction between the nanoparticles leads
to collective plasmons, which are delocalized over the metacrystal. The strong
coupling of these collective plasmons to photons results in the formation of
plasmon polaritons.

Our model readily incorporates retardation effects and considers the
dielectric properties of the nanoparticles and of the medium in which they are
embedded. This has enabled us to derive semi-analytical expressions, which
  determine collective plasmon dispersions, plasmon-polariton dispersions and
  their corresponding polarization dependence, and we have analyzed these
  aspects in detail for the three cubic lattices.  We have discussed the
influence of the dielectric screening due to core electrons of the
nanoparticles and due to the embedding medium on these optical
properties. Specifically, we have shown that the polaritonic dispersions
present band splittings in the near-infrared to the visible range of the
spectrum for all three cubic lattices and for all high-symmetry axes starting
from the center of the first Brillouin zone. Remarkably, for special
directions in the reciprocal space the polaritonic splitting depends on the
polarization, suggesting the possibility to realize a birefringent
metacrystal, despite the high degree of cubic symmetry of the latter. By
comparing our model to classical electrodynamics simulations, we have shown
that it is in quantitative agreement at much reduced computational costs. This
robustness emphasizes that the predicted polarization-dependent band
  dispersions and band splittings should be observable.

\begin{acknowledgments}
We thank Pierre Gilliot for enlightening discussions and Charles A.\ Downing for his
careful reading of the manuscript.  S.L.\ and F.P.\ acknowledge funding through the
Junior Professorship Program of the Ministry of Science, Research and the Arts
(MWK) of Baden-W\"urttemberg within the project ``Theory of Plasmonic
Nanostructures'', through the Carl Zeiss Foundation and the Collaborative
Research Center (SFB) 767 of the German Research Foundation (DFG). 
C.-R.M.\ would like to acknowledge financial support from the EPSRC Center for Doctoral Training in Metamaterials (Grant No.\ EP/L015331/1).
C.-R.M.\ and E.M.\
acknowledge financial support by the Royal Society (International Exchange
Grant No.~IE140367, Newton Mobility Grant NI160073, Theo Murphy
Award TM160190) and by the Leverhulme Trust (Research Project Grant
RPG-2015-101).  G.W.\ is grateful to the French National Research Agency ANR
(Project No.~ANR-14-CE26-0005 Q-MetaMat) and the CNRS PICS program (Contract
No.~6384 APAG) for financial support.  Part of this work was performed on the
computational resource bwUniCluster, funded by the MWK and the universities of
the state of Baden-W\"urttemberg within the framework program bwHPC.
\end{acknowledgments}

\setcounter{equation}{0}
\renewcommand{\theequation}{A\arabic{equation}}
\section*{Appendix: Dielectric tensor}
\label{sec:dielectric}
In this Appendix, we show that our model of interacting plasmonic
nanoparticles leads to a nonlocal, dispersive response. The dielectric tensor
of the metamaterial is calculated explicity for a special, analytically
tractable case, and is found to depend on both the wavevector and the
frequency.

We consider the sc crystal and assume $\bsl{q}=q \hat x$. In this case, the
matrix $F_\bsl{q}$ is diagonal and $f_{\bsl{q}}^{\hat x,\hat x}\neq
f_{\bsl{q}}^{\hat y,\hat y}=f_{\bsl{q}}^{\hat z,\hat z}$. Furthermore, the
choice of $\bsl{q}$ results in a sparse matrix $P_\bsl{q}$ with the only
nonvanishing components being $P_{\bsl{q}}^{\hat y,\hat \lambda_{1, \bsl{q}}}
= P_{\bsl{q}}^{\hat z,\hat \lambda_{2, \bsl{q}}}$.  Hence the matrix on the
left-hand side of Eq.~\eqref{eq:eigensystem}, which we now call $M_{\bsl{q}}$,
can be reordered into a block-diagonal form with block matrices
$M_{x,\bsl{q}}\neq M_{y,\bsl{q}}=M_{z,\bsl{q}}$, which read
  \begin{equation}
    M_{x,\bsl{q}} = \begin{pmatrix} \omega_0 +2\Omega f_{\bsl{q}}^{\hat x,\hat
        x} & -2\Omega f_{\bsl{q}}^{\hat x,\hat x} \\[.1cm]
      2\Omega f_{\bsl{q}}^{\hat x,\hat x} & -\omega_0 -2\Omega f_{\bsl{q}}^{\hat x,\hat x}
      \end{pmatrix}
\end{equation}
and
\begin{widetext}
  \begin{equation}
    M_{y,\bsl{q}} = \begin{pmatrix} \omega_0+2\Omega
      f_{\bsl{q}}^{\hat y,\hat y} & -2\Omega f_{\bsl{q}}^{\hat y,\hat y} & -\mathrm{i}\omega_0\xi_\bsl{q} &
      \mathrm{i}\omega_0\xi_\bsl{q} \\[.1cm] 2\Omega
      f_{\bsl{q}}^{\hat y,\hat y} & -\omega_0 -2\Omega
      f_{\bsl{q}}^{\hat y,\hat y} & \mathrm{i}\omega_0\xi_\bsl{q} &
      -\mathrm{i}\omega_0\xi_\bsl{q} \\[.1cm] \mathrm{i}\omega_0\xi_\bsl{q} &
      \mathrm{i}\omega_0\xi_\bsl{q} & \omega_{\mathrm{ph},
          \bsl{q}}+2\omega_0\xi_\bsl{q}^2 & -2\omega_0\xi_\bsl{q}^2 \\[.1cm]
      \mathrm{i}\omega_0\xi_\bsl{q} & \mathrm{i}\omega_0\xi_\bsl{q} & 2\omega_0\xi_\bsl{q}^2 &
      -\omega_{\mathrm{ph}, \bsl{q}}-2\omega_0\xi_\bsl{q}^2.\\
      \end{pmatrix}.
  \end{equation}
\end{widetext}
The matrix $M_{x,\bsl{q}}$ leads to the longitudinal plasmon, which does not
couple to light within our model. For this reason, we concentrate on the
transverse components. We follow Hopfield \cite{hopfield} to find an
expression for the transverse components of the dielectric tensor of the
metamaterial $\epsilon_{\mathrm{meta}}^{\hat y,\hat
  y}(\bsl{q},\omega)=\epsilon_{\mathrm{meta}}^{\hat z,\hat
  z}(\bsl{q},\omega)$. For this purpose, we calculate
$\det\left(M_{y,\bsl{q}}-\omega^2\mathbbm{1}_{4} \right)=0$ and substitute the
definition of the dielectric function $c^2q^2= \epsilon_{\mathrm{meta}}^{\hat
  y\hat y}(\bsl{q},\omega) \omega^2$ in the resulting expressions. Solving for
$\epsilon_{\mathrm{meta}}^{\hat y\hat y}(\bsl{q},\omega)$ and exploiting the
plasmonic dispersion relation $(\omega_{\mathrm{pl},\bsl{q}}^{\hat
  y})^2=\omega_0^2+4 \Omega \omega_0 f_{\bsl{q}}^{\hat y,\hat y}$
finally yields
\begin{equation}
  \epsilon_{\mathrm{meta}}^{\hat y\hat y}(\bsl{q},\omega) =
  \epsilon_{\mathrm{m}} \left[ 1 + \frac{8 \pi \Omega
    \omega_0}{(\omega_{\mathrm{pl},\bsl{q}}^{\hat y})^2-\omega
    ^2}\right].\label{eq:epsyy}
\end{equation}
Equation~\eqref{eq:epsyy} is the same expression as Eq.~(21) in
Ref.~\cite{Weick} for $\epsilon_{\mathrm{m}}=\epsilon_{\mathrm{d}}=1$ and for
the respective polarization of the collective plasmon, but we consider here 
the dipole-dipole interaction beyond the nearest-neighbor limit. 

\bibliography{refs_cassi.bib}

\end{document}